\documentclass[a4paper]{jpconf}
\usepackage{graphicx}
\newcommand{\Eqn}[1]{&\hspace{-0.5em}#1\hspace{-0.5em}&}

\begin{document}
\title{Effects of ferromagnetic fluctuations on the electric and thermal
transport properties in Na$_x$CoO$_2$ }

\author{Y. Yanagi and Y. \=Ono}

\address{Department of Physics, Niigata University, Niigata, Japan}

\ead{yanagi@phys.sc.niigata-u.ac.jp}

\begin{abstract}
 We investigate the electronic states and the transport properties of
 the CoO$_2$ plane in Na$_x$CoO$_2$ on
 the basis of the two-dimensional triangular lattice 11-band $d$-$p$ model by using the
 fluctuation exchange approximation, where we consider the Coulomb interaction
 between the $t_{2g}$ electrons on a Co site. It is found that all of the
 effective mass of quasiparticles, the resistivity, the
 thermoelectric power and the uniform spin susceptibility increase with
 increasing $x$ for $x>0.6$. This implies that the ferromagnetic fluctuations play
  significant roles on determining the electronic states. These results are qualitatively consistent with the experiments.
\end{abstract}

\section{Introduction}
 The layered cobalt oxide Na$_x$CoO$_2$ has been intensively investigated
 as a promising thermoelectric material due to the large
 thermoelectric power (TEP) together with the low resistivity for
 $x>0.7$\cite{terasaki}. In addition to the large TEP, this compound
 shows rich physical properties. For $x\geq0.75$, a weak
 magnetic order appear below $T_m=22$K\cite{motohashi}, where
 the ferromagnetic ordered CoO$_2$ planes couple antiferromagnetically
 with each other. For $0.6<x<0.75$, the anomalous metallic behavior is
 observed, where the temperature dependence of the uniform spin
 susceptibility is Curie-Weiss-like\cite{yokoi,foo} and the
 thermoelectric power $S$,
 the resistivity $\rho$ and the electronic specific coefficient $\gamma$
 increase with increasing $x$ towards the magnetic ordered
 region\cite{yokoi,lee}. 
 
 Some theoretical studies on Na$_x$CoO$_2$ in Na-rich region have been
 performed.  For example,  Koshibae \textit{et
 al}. studied the TEP in the atomic limit by using the Heikes formula focusing on the
 large spin-orbital
 degeneracy of Co$^{3+}$ and Co$^{4+}$\cite{koshibae}. Effects of the
 band
 structure, however,  
 are not considered there. Kuroki \textit{et al}. investigated the transport properties
 based on the single band Hubbard model and found that the peculiar
 shape of the band structure called ``pudding mold'' plays important
 roles on the TEP together with the low resistivity. Effects of the
 electron correlation, however, are not explicitly  taken into
 account\cite{kuroki}.  Therefore, the microscopic theory
 including both effects of the band structure and the electron correlation is highly desired.

 In the present study, we investigate the electronic states and the transport
 properties of the CoO$_2$ plane in Na$_x$CoO$_2$ by using the
 multi-orbital formula of the fluctuation exchange (FLEX) approximation\cite{takimoto,yada}
 on the basis of the two-dimensional triangular lattice 11-band $d$-$p$
 model\cite{yada,ono,yamakawa,yanagi}. 
\section{Model and Formulation}
Our model includes $3d$ orbitals
($d_{3z^2-r^2},d_{x^2-y^2},d_{xy},d_{yz},d_{zx}$) of Co atoms on the
two-dimensional triangular lattice and $2p$
orbitals ($p_x,p_y,p_z$) of O atoms which lie on the upper and the
lower plane of the Co plane, and is given by the following Hamiltonian
\begin{eqnarray}
H\Eqn{=}\sum_{i,\ell,\sigma}\hspace{-1mm}
\varepsilon^d_{\ell\ell'}d^{\dag}_{i\ell\sigma}d_{i\ell\sigma}
+\hspace{-1mm}\sum_{i,m,\sigma}\hspace{-1mm}\varepsilon^p_{m}p^{\dag}_{im\sigma}p_{im\sigma}
+\sum_{i,j,\ell,\ell',\sigma}\hspace{-1mm}t^{dd}_{i,j,\ell,\ell'}d^{\dag}_{i\ell\sigma}d_{j\ell'\sigma}\nonumber\\
\Eqn{+}\hspace{-1mm} \sum_{i,j,m,m',\sigma}\hspace{-1mm}t^{pp}_{i,j,m,m'}p^{\dag}_{im\sigma}p_{jm'\sigma}
+\sum_{i,j,\ell,m,\sigma}\hspace{-1mm}t^{dp}_{i,j,\ell,m}d^{\dag}_{i\ell\sigma}p_{jm\sigma}+h.c.\nonumber\\
\Eqn{+} H_{\mathrm{int}} \label{d-p}, 
\end{eqnarray}
where $d_{i\ell\sigma}$ is the annihilation operator for a Co-$3d$ electron with spin
$\sigma$ in the  orbital $\ell$ at  site $i$ and $p_{im\sigma}$ is the annihilation
operator for a O-$2p$ electron with spin
$\sigma$ in the  orbital $m$ at  site $i$. 
In eq. (\ref{d-p}), the
transfer integrals $t^{dd}_{i,j,\ell,\ell'}$, $t^{pp}_{i,j,m,m'}$,
$t^{dp}_{i,j,\ell,m}$ and the atomic energies $\varepsilon^d_{\ell\ell'}$,
$\varepsilon^p_{m}$ are determined so as to fit the energy bands obtained from the
tight-binding approximation to
those from the band calculation\cite{singh} and are listed
in ref. 9. The band calculation predicted the large hole Fermi
surface around the $\Gamma$-point and the six small hole pockets near
the $K$-points\cite{singh}. However, the six small hole pockets have not been
observed in the ARPES measurements\cite{yang}. Thus, we introduce the change in the crystal field energy 
$\sum_{i,\ell\neq\ell'\in t_{2g},\sigma}V_t'd^{\dag}_{i\ell\sigma}d_{i\ell'\sigma}$
 and then set $V_t'=0.15$eV  to sink those below the
 Fermi level in the present study\cite{yada}.
 $H_{\mathrm{int}}$ is the Hamiltonian of the Coulomb interaction between the $t_{2g}$ electrons on the Co atom and characterized by the following parameters: 
the intra (inter)-orbital direct terms $U$ ($U'$), the Hund's rule
coupling $J$ and
the pair transfer $J'$\cite{yada}.

In the FLEX approximation, the spin (charge-orbital) susceptibility
$\hat{\chi}^s(q)$ ($\hat{\chi}^c(q)$) is given as follows
\begin{equation}
\hat{\chi}^s(q)=(\hat{1}-\hat{\chi}^{(0)}(q)\hat{S})^{-1}\hat{\chi}^{(0)}(q), \quad
\hat{\chi}^c(q)=(\hat{1}+\hat{\chi}^{(0)}(q)\hat{C})^{-1}\hat{\chi}^{(0)}(q),
\end{equation}
where $q=(\mathbf{q},i\omega_m=i2m\pi T)$.  In the above, $\hat{\chi}^{(0)}(q)$ 
 and $\hat{S}(\hat{C})$ are the matrices whose components are given by
\begin{equation}
\chi^{(0)}_{\ell_1\ell_2,\ell_3\ell_4}(q)=-\frac{T}{N}\sum_{k}G_{\ell_3\ell_1}(k)G_{\ell_2\ell_4}(k+q),
\end{equation}
\begin{equation}
S_{\ell_1\ell_2,\ell_3\ell_4}~(C_{\ell_1\ell_2,\ell_3\ell_4})= \left\{
\begin{array}{@{\,} l @{\,} c}
U~(U) & (\ell_1=\ell_2=\ell_3=\ell_4)\\
U'~(-U'+2J) & (\ell_1=\ell_3\ne\ell_2=\ell_4)\\
J~(2U'-J) & (\ell_1=\ell_2\ne\ell_3=\ell_4)\\
J'~(J')& (\ell_1=\ell_4\ne\ell_2=\ell_3)\\
0 & (\mathrm{otherwise})
\end{array} \right. ,\nonumber
\end{equation} 
where $k=(\mathbf{k},i\varepsilon_n=i(2n+1)\pi T)$,
$\hat{G}(k)=(\{\hat{G}^{(0)}(k)\}^{-1}-\hat{\Sigma}(k))^{-1}$ and
$\hat{G}^{(0)}(k)$ is the noninteracting Green's function. 
 The self-energy $\hat{\Sigma}(k)$ is
given by
\begin{equation}
\Sigma_{\ell\ell'}(q)=\frac{T}{N}\sum_{q,\ell_1,\ell_2}G_{\ell_1\ell_2}(k-q)V^{\mathrm{eff}}_{\ell_1\ell,\ell_2\ell'}(q),
\end{equation}
where the effective interaction is
$\hat{V}^{\mathrm{eff}}(q)=\frac{3}{2}\hat{V}^s(q)+\frac{1}{2}\hat{V}^c(q)$
with  $\hat{V}^s(q)=\hat{S}\hat{\chi}^s(q)\hat{S}-\frac{1}{2}\hat{S}\hat{\chi}^{(0)}(q)\hat{S}$
 and $\hat{V}^c(q)=\hat{C}\hat{\chi}^c(q)\hat{C}-\frac{1}{2}\hat{C}\hat{\chi}^{(0)}(q)\hat{C}.$
We solve the above equations (2)-(5) self-consistently to obtain
$\hat{G}(k), \hat{\chi}^s(q)$ and $\hat{\chi}^c(q)$, and then calculate $\rho$ and $S$ based on
the linear response theory, where we neglect the vertex corrections. For simplicity, we assume $U'=U-2J$, $J=J'$ and set $U=1.5$eV,
$J=0.15$eV in the present study. In the numerical calculation, we use the $64\times64$ $\mathbf{k}$-points and 256
Matsubara frequencies.

\section{Results}
\begin{figure}[t]
\begin{center}
\begin{minipage}{7.5cm}
\includegraphics[width=7.5cm]{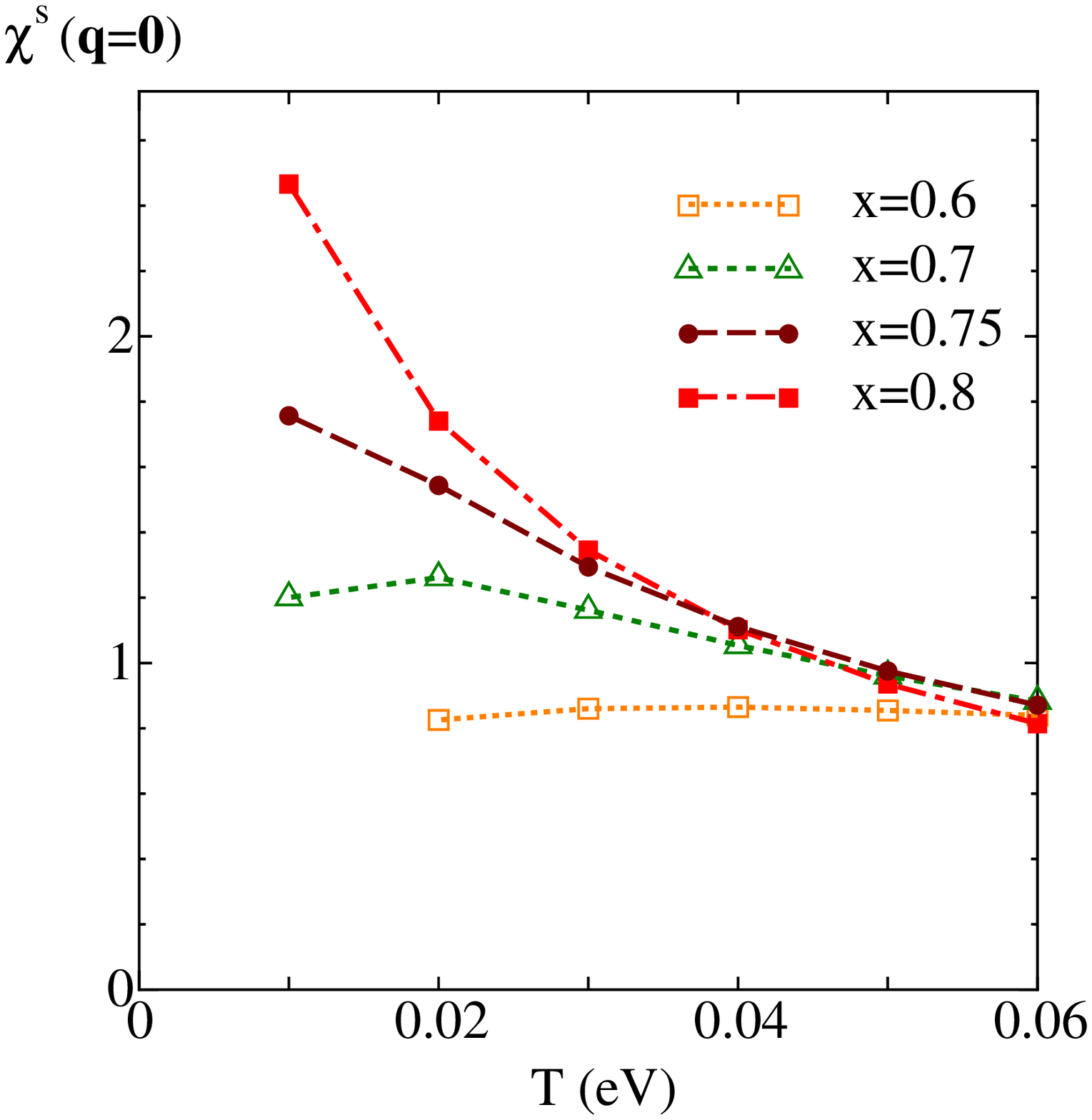}
\caption{\label{chis}Temperature dependence of the uniform spin
 susceptibility $\chi^s$.}
\end{minipage}\hspace{5mm}%
\begin{minipage}{6.5cm}
\includegraphics[width=6.5cm]{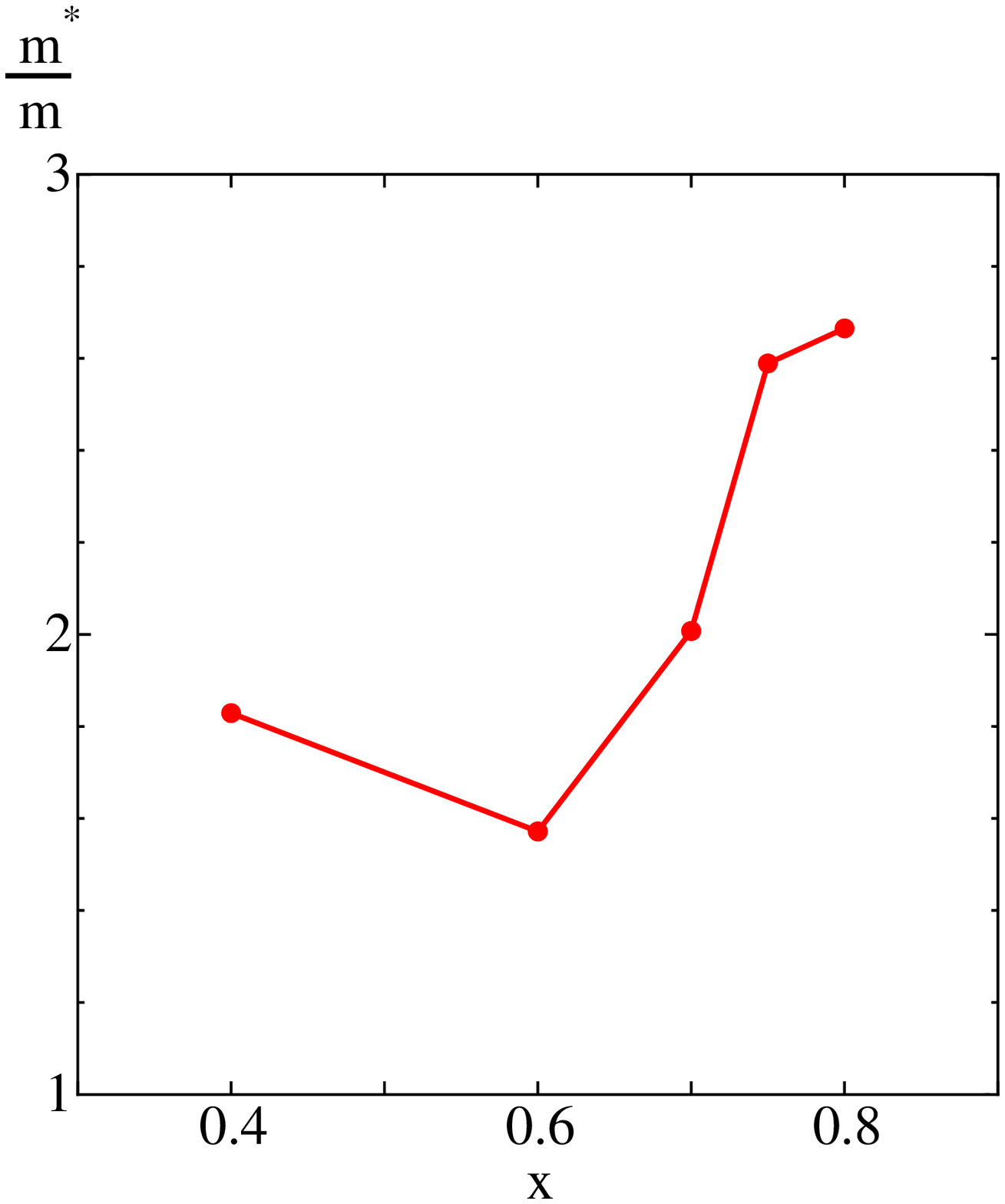}
\caption{\label{mass}Doping dependence of the effective mass of  quasiparticles $m^*/m$}
\end{minipage}
\end{center} 
\end{figure}

We show the temperature dependence of the uniform spin susceptibility
$\chi^s=\sum_{\ell\ell'}\chi^s_{\ell\ell,\ell'\ell'}(\mathbf{q}=\mathbf{0},i\omega_m=0)$
in Figure \ref{chis}.  For $x\geq0.75$, $\chi^s$ shows the
Curie-Weiss-like behavior, while for $x\leq0.7$, it shows the weak
pseudogap like behavior at low temperatures. 
At the low temperature, $\chi^s$ increases with the doping $x$ due to the
effect of  the flat dispersion near the $\Gamma$-point ($\mathbf{k}$=(0,0)) which approaches the Fermi
level with increasing $x$.

 Figure 2 shows the doping dependence of the effective mass
of quasiparticles,
 $m^*/m=z_{k_F}^{-1}=1-\left.\frac{\partial\mathrm{Re}\Sigma^{\mathrm{R}}_{\alpha}(k_F,\varepsilon)}{\partial\varepsilon}\right|_{\varepsilon=0}$, 
 where $\Sigma_{\alpha}^{\mathrm{R}}(\mathbf{k},\varepsilon)$ is the retarded self-energy in
the band basis which is obtained by the unitary transformation and the numerical analytic continuation
of $\Sigma_{\ell\ell'}(\mathbf{k},i\varepsilon_n)$ from the upper half plane to the real
axis with use of the Pad\'e approximation. In the present paper, all of the
numerical calculations are performed for $T\geq0.01\mathrm{eV}$. Then, we
obtain the extrapolated value of $z_{k_F}$ at $T=0\mathrm{eV}$ from the data at
$T=0.01-0.06\mathrm{eV}$ by using the third order polynomial fit.  For $x\geq0.6$, $m^*/m$ increases with increasing
$x$, as observed in experiments\cite{yokoi}. The increase in $m^*/m$ with the
doping $x$ is considered due to the effect of the ferromagnetic
fluctuations as shown in $\chi^s$ at the low temperature.

Finally, we turn to the transport properties. Figure 3. (a) shows the temperature dependence
of $\rho$.  We find that $\rho$ increases with increasing
$x$, and shows concave downward temperature dependence for $x\geq0.7$.
Figure 3. (b) shows  the temperature dependence of $S$. 
$S$ is always positive and increases with
increasing $x$. The increases in  both $\rho$ and
$S$ with the doping $x$ are qualitatively consistent with
 experiments\cite{lee}.  
\begin{figure}[h]
\begin{center}
\begin{minipage}{6.5cm}
\includegraphics[width=6.5cm]{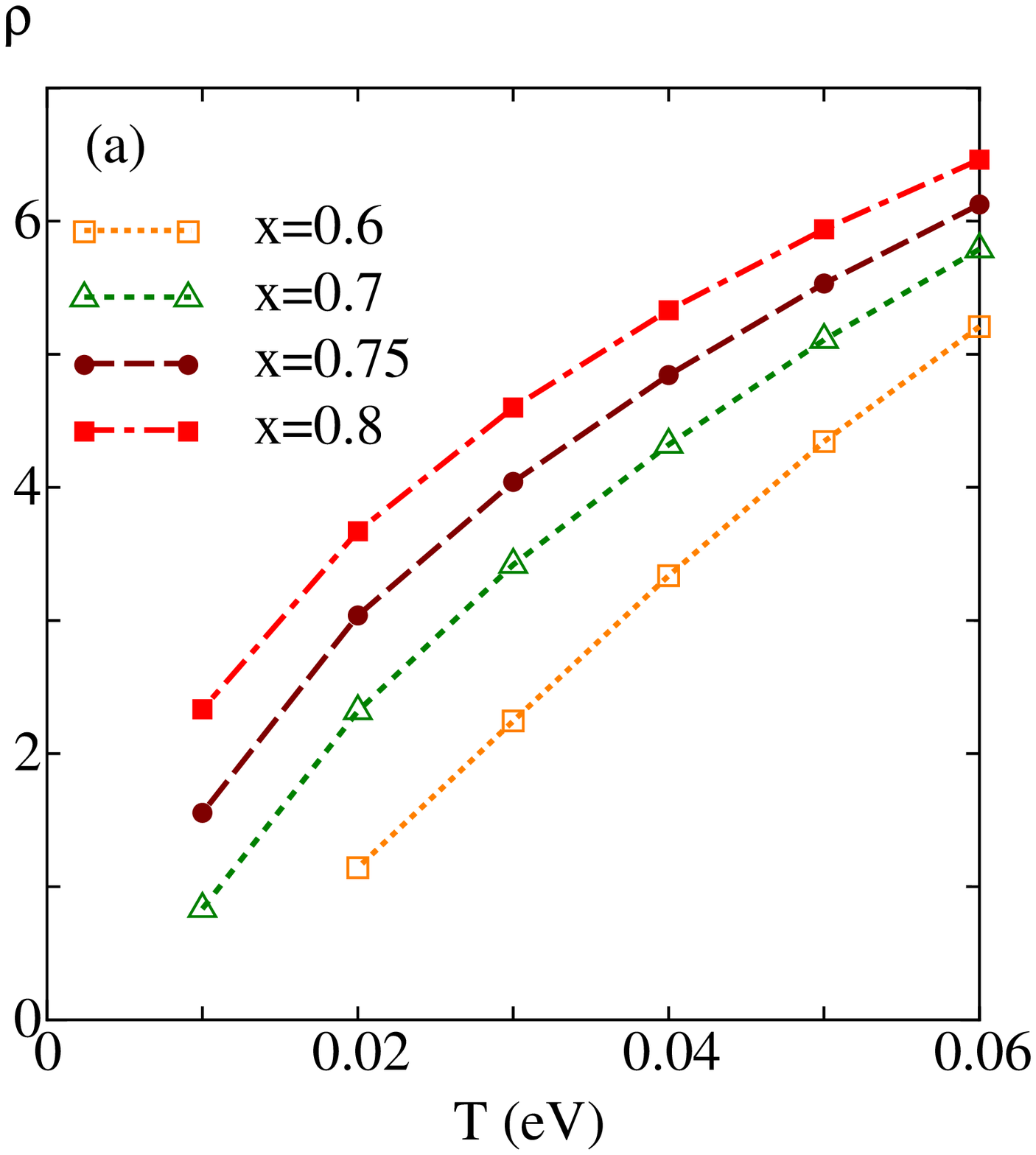}
\end{minipage}
\begin{minipage}{7.0cm}
\includegraphics[width=7.0cm]{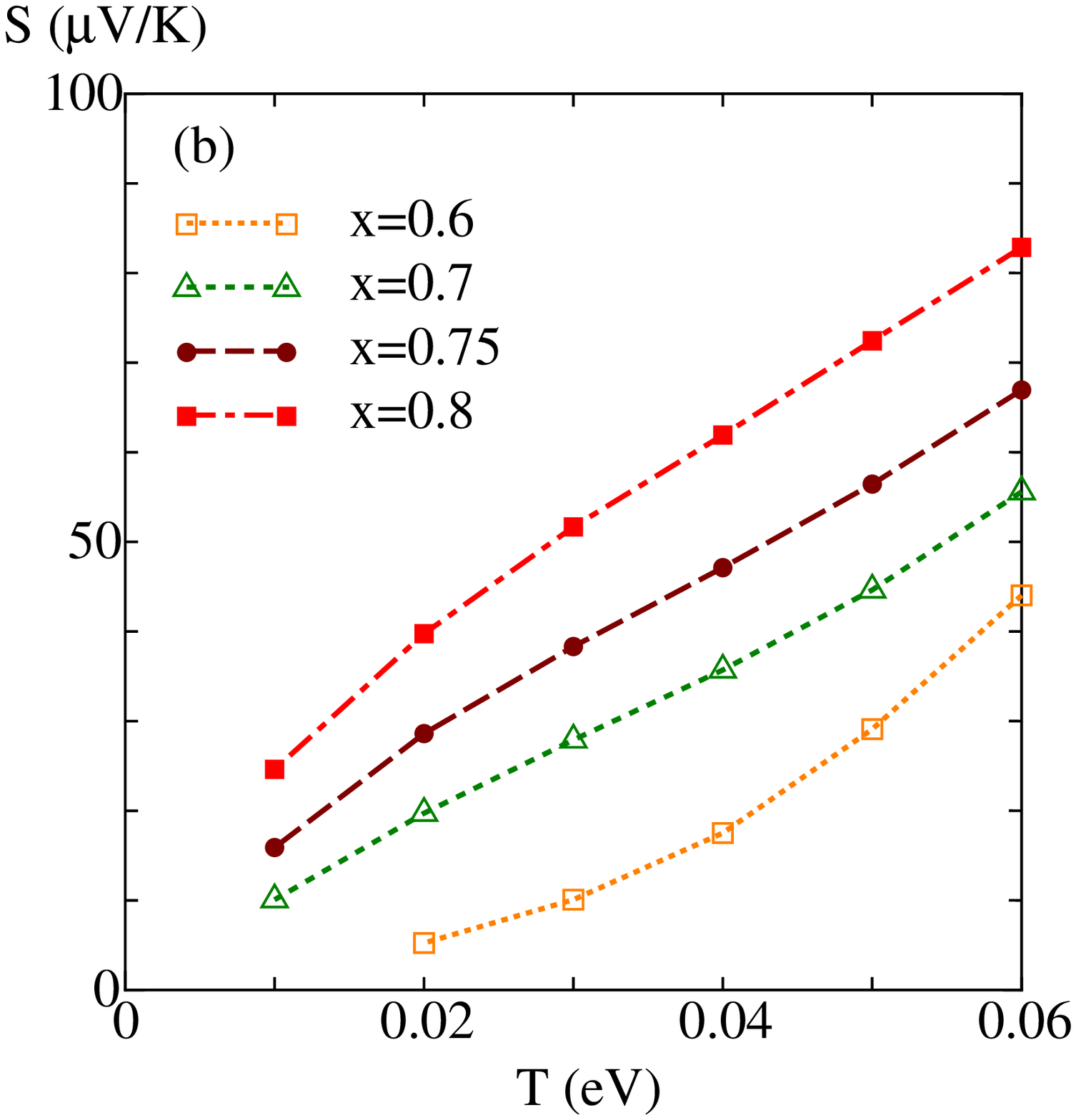}
\end{minipage}
\caption{\label{trans} Temperature dependence of (a) the resistivity
 $\rho$ and (b) the thermoelectric power $S$.} 
\end{center}
\end{figure}
\section{Summary and Discussion}
We have investigated the electronic states and the transport properties of
the CoO$_2$ plane in Na$_x$CoO$_2$ on the basis of the two-dimensional
triangular lattice 11-band $d$-$p$ model by using the FLEX
approximation. It has been found that the temperature dependence of $\chi^s$
shows the Curie-Weiss-like behavior for $x\geq0.75$ and $m^*/m$, $\rho$,
$S$ and $\chi^s$ increase with increasing $x$ for $x\geq0.6$ due to the effects of the ferromagnetic fluctuations. These results are qualitatively consistent with  experimental results\cite{yokoi,foo,lee}. 

However, we failed to reproduce the hump in the temperature dependence of $S$ experimentally observed at $T\sim50-150\mathrm{K}$\cite{lee}. 
In addition, the absolute value of $S$ obtained from our calculation is less than a half of that from experiments\cite{terasaki,lee}. These results seem to be caused by underestimation of the ferromagnetic fluctuations in the FLEX approximation. The effects of the mode-mode coupling due to the vertex corrections, which are considered to play crucial roles for the nearly ferromagnetic metals as shown in the self-consistent renormalization theory\cite{moriya}, are neglected in the FLEX approximation and will be discussed in the subsequent paper. 

\ack
The authors thank H. Kontani and K. Yada for many useful comments and
discussions.

\end{document}